\documentclass[12pt]{article}
\usepackage[latin1]{inputenc}
\usepackage{pstricks}
\usepackage{amsmath}
\usepackage{float}
\usepackage{color}
\input epsf
\usepackage{amssymb}
\usepackage[dvips]{graphicx,psfrag}
\newcommand{\be}{\begin{equation}}
\newcommand{\ee}{\end{equation}}
\newcommand{\bea}{\begin{eqnarray}}
\newcommand{\eea}{\end{eqnarray}}
\newcommand{\lb}{\label}
\usepackage{graphicx}
\usepackage{amssymb}
\usepackage{epstopdf}
\def \g{\gamma}

\begin{document}
\begin{titlepage}
\title{Global properties of the growth index of matter inhomogeneities in the Universe}

\author{
R. Calderon$^1$\thanks{email:rodrigo.calderon-bruni@umontpellier.fr}, 
D. Felbacq$^1$\thanks{email:didier.felbacq@umontpellier.fr}, 
R. Gannouji$^2$\thanks{email:radouane.gannouji@pucv.cl}, 
D. Polarski$^1$\thanks{email:david.polarski@umontpellier.fr},
A.A.~Starobinsky$^{3,4}$\thanks{email:alstar@landau.ac.ru}
\hfill\\
$^1$ Laboratoire Charles Coulomb, Universit\'e Montpellier \& CNRS\\
 UMR 5221, F-34095 Montpellier, France\\
$^2$ Instituto de F\'{\i}sica, Pontificia Universidad  Cat\'olica de Valpara\'{\i}so,\\
 Casilla 4950, Valpara\'{\i}so, Chile\\
$^3$ Landau Institute for Theoretical Physics RAS, Moscow, 119334, Russia\\
$^4$ National Research University Higher School of Economics,\\
 Moscow 101000, Russia }
\pagestyle{plain}
\date{\today}

\maketitle

\begin{abstract}
We perform here a global analysis of the growth index $\gamma$ behaviour from deep in the 
matter era till the far future. For a given cosmological model in GR or in modified gravity, 
the value of $\gamma(\Omega_m)$ is unique when the decaying mode of scalar perturbations is 
negligible. However, $\g_{\infty}$, the value of $\gamma$ in the asymptotic future, is 
unique even in the presence of a nonnegligible decaying mode today. Moreover $\g$ becomes 
arbitrarily large deep in the matter era. Only in the limit of a vanishing decaying mode do 
we get a finite $\g$, from the past to the future in this case. We find further a condition 
for $\g(\Omega_m)$ to be monotonically decreasing (or increasing). This condition can be 
violated inside general relativity (GR) for varying $w_{DE}$ though generically $\g(\Omega_{m})$ 
will 
be monotonically decreasing (like $\Lambda$CDM), except in the far future and past. A bump or a 
dip in $G_{\rm eff}$ can also lead to a significant and rapid change in the slope 
$\frac{d\g}{d\Omega_{m}}$. 
On a $\Lambda$CDM background, a $\g$ substantially lower (higher) 
than $0.55$ with a negative (positive) slope reflects the opposite evolution of $G_{\rm eff}$. 
In DGP models, $\g(\Omega_{m})$ is monotonically increasing 
%$\frac{d\g}{d\o}>0$ 
except in the far future. 
While DGP gravity becomes weaker than GR in the future and $w^{DGP}\to -1$, we still get  
$\g_{\infty}^{DGP}= \g_{\infty}^{\Lambda CDM}=\frac{2}{3}$. In contrast, despite 
$G^{DGP}_{\rm eff}\to G$ in the past, $\g$ does not tend to its value in GR because 
$\frac{dG^{DGP}_{\rm eff}}{d\Omega_{m}}\Big|_{-\infty}\ne 0$. 
\end{abstract}

%PACS Numbers: 98.80.-k, 95.36.+x
\end{titlepage}

%%%%%%%%%%%%%%%%%%%%%%%%%%%%%%%%%%%%%%%%%%%%%%%%%%%%%%%%%%%%%%%%%%%%%%%%%%%%%%%%%%%%%%%%%%%
%%%%%%%%%%%%%%%%%%%%%%%%%%%%%%%%%%%%%%%%%%%%%%%%%%%%%%%%%%%%%%%%%%%%%%%%%%%%%%%%%%%%%%%%%%%
\section{Introduction}
%%%%%%%%%%%%%%%%%%%%%%%%%%%%%%%%%%%%%%%%%%%%%%%%%%%%%%%%%%%%%%%%%%%%%%%%%%%%%%%%%%%%%%%%%%%
%%%%%%%%%%%%%%%%%%%%%%%%%%%%%%%%%%%%%%%%%%%%%%%%%%%%%%%%%%%%%%%%%%%%%%%%%%%%%%%%%%%%%%%%%%%
Despite intensive activity in recent years, the late-time accelerated expansion rate of 
the Universe remains a theoretical challenge.
A wealth of theoretical models and mechanisms were put forward for its solution, see the 
reviews \cite{SS00}. 
Remarkably, the simplest model -- GR with a non-relativistic matter component (mainly 
a non-baryonic one) and a cosmological constant $\Lambda$ -- is in fair agreement with 
observational data, especially on large cosmic scales. Notwithstanding the theoretical 
problems it raises, 
this model provides a benchmark for the assessment of other dark energy (DE) models 
phenomenology. One can make progress by exploring carefully the phenomenology 
of the proposed models and comparing it with observations \cite{WMEHRR13}. 
Tools which can efficiently discriminate between models, or between classes of models 
(e.g. \cite{SSS14}) are then needed. 
The growth index $\gamma$ which provides a representation of the growth of density 
perturbations in dust-like matter, is an example of such phenomenological tool. 
Its use was pioneered long time ago in order to discriminate spatially open from 
spatially flat universes \cite{P84} and then generalized to other cases \cite{LLPR91}. It 
was later revived in the context of dark energy \cite{LC07}, with the additional incentive to 
single out DE models beyond GR. 
The growth index has a clear and important signature in the presence of $\Lambda$: it 
approaches 6/11 for $z\gg 1$ and it changes little from this value even up to $z=0$. 
This behaviour still holds for smooth noninteracting DE models inside GR with a constant 
equation of state $w_{DE}$. A strictly constant $\gamma$ is very peculiar 
\cite{PG07,PSG16}. 
On the other hand, this behaviour is strongly violated for some models beyond GR, 
see e.g. \cite{GMP08,MSY10}. 
Surprisingly, important properties of the growth index behaviour can be understood 
by making a connection with a strictly constant $\g$. 
Further, while present observations probe low redshifts $z\lesssim 2$, 
still as we will see, more insight is gained looking at the global evolution of $\g$. 
Before starting this analysis, we first review the basic formalism for the study of the growth 
index $\g$.  

%%%%%%%%%%%%%%%%%%%%%%%%%%%%%%%%%%%%%%%%%%%%%%%%%%%%%%%%%%%%%%%%%%%%%%%%%%%%%%%%%%%%%%%%%%%
%%%%%%%%%%%%%%%%%%%%%%%%%%%%%%%%%%%%%%%%%%%%%%%%%%%%%%%%%%%%%%%%%%%%%%%%%%%%%%%%%%%%%%%%%%%
\section{The growth index}
%%%%%%%%%%%%%%%%%%%%%%%%%%%%%%%%%%%%%%%%%%%%%%%%%%%%%%%%%%%%%%%%%%%%%%%%%%%%%%%%%%%%%%%%%%%
%%%%%%%%%%%%%%%%%%%%%%%%%%%%%%%%%%%%%%%%%%%%%%%%%%%%%%%%%%%%%%%%%%%%%%%%%%%%%%%%%%%%%%%%%%%
In this section, we review the basic equations and definitions concerning the growth index 
$\gamma$. 
Let us consider the dynamics on cosmic scales  at the linear level of density 
perturbations $\delta_m =\delta\rho_m/\rho_m$ in the dust-like matter component.  
Deep inside the Hubble radius their evolution obeys the following equation 
\be
{\ddot \delta_m} + 2H {\dot \delta_m} - 4\pi G\rho_m \delta_m = 0~,\label{del}
\ee
where $H(t)\equiv \dot a(t)/a(t)$ stands for the Hubble parameter and $a(t)$ is the 
scale factor of a Friedmann-Lema\^itre-Robertson-Walker (FLRW) Universe filled with 
standard dust-like matter and DE (we neglect radiation at the matter and DE dominated stages). 

For vanishing spatial curvature, we have for $z\ll z_{eq}$
\be
h^2(z) = \Omega_{m,0} ~(1+z)^3 + (1 - \Omega_{m,0})
                      ~\exp \left[ 3\int_{0}^z dz'~\frac{1+w_{DE}(z')}{1+z'}\right]~,\lb{h2z}
\ee 
with $h(z)\equiv \frac{H}{H_0}$, $w_{DE}(z)\equiv p_{DE}(z)/\rho_{DE}(z)$, $z=\frac{a_0}{a}-1$, 
and finally $\Omega_{m,0}\equiv \frac{\rho_{m,0}}{\rho_{{\rm cr},0}}$. 
Equality \eqref{h2z} will hold for all FLRW models inside GR. It is still valid for many 
models beyond GR with appropriate definition of the dark energy sector. 
We recall the definition
\be
\Omega_m=\Omega_{m,0} \frac{a_0^3}{a^3} h^{-2}~, \lb{Om}
\ee
and the useful relation
\be 
w_{DE} = \frac{1}{3(1-\Omega_m)}~ \frac{d\ln \Omega_m}{d\ln a}~.  \label{wDE}
\ee 
Introducing the growth function $f\equiv \frac{d \ln \delta_m}{d \ln a}$ and using 
\eqref{wDE}, equation \eqref{del} leads to the equivalent nonlinear first order 
equation \cite{WS98}
\be
\frac{df}{dN} + f^2 + \frac{1}{2} \left(1 - \frac{d \ln \Omega_m}{dN} \right) f = 
                              \frac{3}{2}~\Omega_m \lb{df}
\ee
with $N\equiv \ln a$. Clearly $f=p$ if $\delta_m\propto a^p$ (with $p$ constant). 
In particular, when the growing mode dominates, $f\to 1$ in $\Lambda$CDM for large $z$ 
and $f=1$ in the Einstein-de Sitter universe. 
In the peculiar situation where the decaying mode dominates, $f\to -\frac32$ in $\Lambda$CDM 
for large $z$ and $f=-\frac32$ in the Einstein-de Sitter universe. We will return later to 
the consequences of a nonnegligible decaying mode. 
Also, if the dependence $f(N)$ is known from observations, the Hubble function $H(z)$ and 
finally the scale factor $a(t)$ dependence can be determined unambiguosly in an analytical 
form \cite{Starobinsky:1998fr} (see \cite{LHuillier:2019imn} for implementation of this to 
recent observational data).

The following parametrization has been intensively used and investigated in the context 
of dark energy
\be
f = \Omega_m(z)^{\gamma(z)}~,\lb{Omgamma}
\ee
where $\gamma$ is dubbed growth index, though in general $\gamma(z)$ is a genuine function. 
It can even depend on scales for models where the growth of matter perturbations has a scale 
dependence. 
The representation (\ref{Omgamma}) has attracted a lot of interest with the aim to 
discriminate between DE models based on modified gravity theories and the $\Lambda$CDM paradigm.
It turns out that the growth index is quasi-constant for the standard $\Lambda$CDM from the 
past untill today. Such a behaviour holds for smooth non-interacting DE models when $w_{DE}$ 
is constant, too \cite{PG07}. 

In many DE models outside GR the modified evolution of matter perturbations is recast into 
\be
{\ddot \delta_m} + 2H {\dot \delta_m} - 4\pi G_{\rm eff}\rho_m\delta_m = 0~,\label{delmod}
\ee
where $G_{\rm eff}$ is some effective gravitational coupling appearing in the model.
For example, for effectively massless scalar-tensor models \cite{BEPS00}, $G_{\rm eff}$ is 
varying with time but it has no scale dependence while its value today is equal to 
the usual Newton's constant $G$. Introducing for convenience the quantity 
\be
g\equiv \frac{G_{\rm eff}}{G}\lb{g}~,
\ee  
equation \eqref{delmod} is easily recast into the modified version of Eq. \eqref{df}, 
viz.
\be
\frac{df}{dN} + f^2 + \frac{1}{2} \left(1 - \frac{d \ln \Omega_m}{dN} \right) f = 
                              \frac{3}{2}~g~\Omega_m~,\lb{dfmod}
\ee
where the same GR definition $\Omega_m = \frac{8\pi G \rho_m}{3 H^2}$ is used. 
Some subtleties can arise if the defined energy density of DE becomes negative. 
When the growth index $\gamma$ is strictly constant, it is straightforward to deduce from 
\eqref{dfmod} that $w_{DE}=\overline{w}$ with
\bea
\overline{w} &=& - \frac{1}{3(2\gamma -1)} ~
\frac{1 + 2\Omega_m^{\gamma} - 3 g \Omega_m^{1-\gamma}}{1-\Omega_m} \lb{wgcon}\\
&\equiv& - \frac{1}{3(2\gamma -1)} ~ F(\Omega_m; g, \gamma). 
\eea
The quantity $\overline{w}(\Omega_m,\gamma)$ defines a function in the 
$(\Omega_m, \gamma)$ plane which will be useful even when $\gamma$ is not constant. 
The case $g=1$ reduces to GR and we will simply write
\be 
F(\Omega_m; g=1, \gamma)\equiv F(\Omega_m; \gamma)~.
\ee
Below, for any quantity $v$, $v_{\infty}$, resp. $v_{-\infty}$, will denote its (limiting) 
value for $N\to \infty$ in the DE dominated era ($\Omega_m\to 0$), resp. 
$N\to -\infty$ (generically $\Omega_m\to 1$). 
We have in particular from \eqref{wgcon} for $g=1$ (GR)
\bea
\gamma &=& \frac{3 \overline{w}_{\infty} - 1}{6 \overline{w}_{\infty}} \lb{gconstasf}\\
\gamma &=& \frac{3(1 - \overline{w}_{-\infty})}{5 - 6 \overline{w}_{-\infty}}~.\lb{gconstasp}
\eea
Here we assume that $\overline{w}<0$ to get matter dominated stage in the
past and dark energy dominated stage in the future. In addition, Eq. (13)
requires that $\overline{w}_{\infty} < -\frac13$ to have $0< \gamma <1$,
otherwise $\overline{w}_{\infty}$ becomes infinite.

As it was found in \cite{PSG16}, these relations between a \emph{constant} $\g$ and the 
corresponding asymptotic values $\overline{w}_{\infty}$ and $\overline{w}_{-\infty}$ apply 
also for the \emph{dynamical} $\g$ obtained for an arbitrary but given $w_{DE}$. 
In the latter case, we obtain for $g=1$  
\bea
\gamma_{\infty} &=& \frac{3 w_{\infty} - 1}{6 w_{\infty}} \lb{gasf}\\
\gamma_{-\infty} &=& \frac{3(1 - w_{-\infty})}{5 - 6 w_{-\infty}}~,\lb{gasp}
\eea
with $w_{\infty}$, resp. $w_{-\infty}$, the asymptotic value of $w_{DE}$ in the future, resp. 
past. We will return later to this important property. 
%%%%%%%%%%%%%%%%%%%%%%%%%%%%%%%%%%%%%%%%%%%%%%%%%%%%%%%%%%%%%%%%%%%%%%%%%%%%%%%%%%%%%%%%%%%%%%
%%%%%%%%%%%%%%%%%%%%%%%%%%%%%%%%%%%%%%%%%%%%%%%%%%%%%%%%%%%%%%%%%%%%%%%%%%%%%%%%%%%%%%%%%%%%%%
%%%%%%%%%%%%%%%%%%%%%%%%%%%%%%%%%%%%%%%%%%%%%%%%%%%%%%%%%%%%%%%%%%%%%%%%%%%%%%%%%%%%%%%%%%%%%%

%%%%%%%%%%%%%%%%%%%%%%%%%%%%%%%%%%%%%%%%%%%%%%%%%%%%%%%%%%%%%%%%%%%%%%%%%%%%%%%%%%%%%%%%%%%%%
%%%%%%%%%%%%%%%%%%%%%%%%%%%%%%%%%%%%%%%%%%%%%%%%%%%%%%%%%%%%%%%%%%%%%%%%%%%%%%%%%%%%%%%%%%%%%
\section{Global analysis in the $(\Omega_m,\g)$ plane}
\begin{figure}[h!]
   \begin{center}
   \includegraphics[height=8cm]{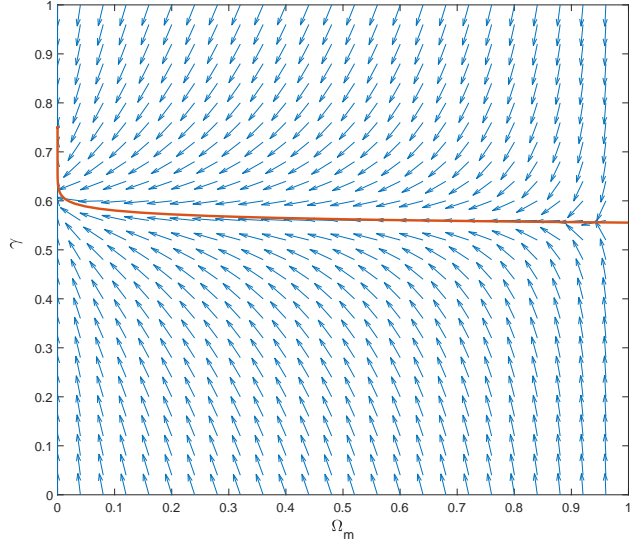}
   \caption{The (unit time-oriented) vector field tangent to the solutions $\g$ of 
   eq.\eqref{dg} is displayed for $w_{DE}=-1$. The red curve corresponds to the solution for 
   $\g$, with boundary condition $\g(1)\equiv \g_{-\infty} = 6/11$. It is the only solution 
   which is finite 
   on the entire $\Omega_m$ interval, physically it corresponds to the presence of the 
   growing mode 
   of \eqref{del} only. The difference can be important in the far past only. All the other 
   solutions correspond to solutions containing also the decaying mode solution. As explained 
   in the text, these solutions (if physical) tend to the same value of $\g$, namely 
   $\gamma_{\infty}=\frac{2}{3}$, for $\Omega_m \to 0$, and they diverge for $\Omega_m \to 1$. 
   Some integral curves $\gamma(\Omega_m)$ are shown on figure \eqref{vecf2}.}
   \label{vecf}
   \end{center}
\end{figure}

%%%%%%%%%%%%%%%%%%%%%%%%%%%%%%%%%%%%%%%%%%%%%%%%%%%%%%%%%%%%%%%%%%%%%%%%%%%%%%%%%%%%%%%%%%%%%
%%%%%%%%%%%%%%%%%%%%%%%%%%%%%%%%%%%%%%%%%%%%%%%%%%%%%%%%%%%%%%%%%%%%%%%%%%%%%%%%%%%%%%%%%%%%%
It is natural to consider $\Omega_m$ as the fundamental integration variable.  
In this case the evolution equation for $\gamma$ obtained from \eqref{dfmod} using 
\eqref{Omgamma} yields
\begin{equation}
2\alpha \Omega_m \ln(\Omega_m) \frac{d \g}{d\Omega_m}+\alpha(2\g-1)+F(\Omega_m;g,\g)=0~, 
\label{dg}
\end{equation}
where we have defined 
\be
\alpha\equiv 3 w_{DE}~.
\ee
We restrict ourselves in this section to GR i.e. $g=1$.
This equation is well defined provided $w_{DE}$ is a given function of $\Omega_m$ only. In 
particular 
\eqref{dg} can be readily used for constant $w_{DE}$. For arbitrary $w_{DE}(a)$ 
we can either express $a$ as a function of $\Omega_m$ or write \eqref{dg} using the 
integration variable $a$.
Note that the function $a(\Omega_m)$ is one-to-one for $w_{DE}<0$.
We consider here all the solutions to equation \eqref{dg} on the entire $\Omega_m$ interval.  
Each integral curve of equation \eqref{dg} is the envelope of its tangent vectors 
\be
\left( 
\begin{array}{c}
1 \\
\frac{d \g}{d\Omega_m}
\end{array}
\right)
\ee
The collection of all these tangent vectors defines a vector field that can be written
\be
\left( 
\begin{array}{c}
2\alpha \Omega_m (1-\Omega_m) \ln(\Omega_m) \\
-\alpha(2\g-1)(1-\Omega_m)-\tilde{F}(\Omega_m;\g)
\end{array}
\right)
\ee
where we have defined 
\be
\tilde{F}(\Omega_m;\g)\equiv (1-\Omega_m)F(\Omega_m;\g)= 1 + 2\Omega_m^{\g} - 
                                   3\Omega_m^{1-\g}~. \lb{tildeF}
\ee
For convenience, we have written the vector field in this way in order to have 
explicitly regular functions, so this vector field is well-defined 
everywhere for $(\Omega_m,\g) \in [0,1] \times \mathbb{R}$.
The associated unit vector field is represented in fig.\eqref{vecf} for $w_{DE}=-1$. 

%%%%%%%%%%%%%%%%%%%%%%%%%%%%%%%%%%%%%%%%%%%%%%%%%%%%%%%%
%%%%%%%%%%%%%%%%%%%%%%%%%%%%%%%%%%%%%%%%%%%%%%%%%%%%%%%
 
The integral curves of this vector field (i.e. the phase portrait), are obtained by solving 
the autonomous differential system
%\be
%\begin{array}{l}
\bea
\frac{d\Omega_m}{ds} &=& 2\alpha \Omega_m (1-\Omega_m) \ln(\Omega_m) \nonumber\\
\frac{d\g}{ds} &=& -\alpha(2\g-1)(1-\Omega_m)-\tilde{F}(\Omega_m;\g)
\eea
where $s \in \mathbb{R}^+$ is a dummy variable parametrizing the curves.

The vector field corresponding to $w_{DE}=-1$ is displayed on fig.\eqref{vecf} and will be 
considered now for concreteness. 
The growth index $\gamma(\Omega_m)$ which starts at $\g(1)\equiv \g_{-\infty} = 6/11$ is the 
only curve $\gamma(\Omega_m)$ which is finite everywhere. This curve corresponds to the  
presence solely of the growing mode of eq.\eqref{del} or equivalently to the limit of a 
vanishing decaying mode. Let us consider this point in more details. Generally, in the 
presence of two modes $\delta \equiv \delta_1 + \delta_2$, we have 
\be
f = \frac{\delta_1}{\delta} f_1 + \frac{\delta_2}{\delta} f_2~,\lb{f12}
\ee
where $f_1>0$, resp. $f_2<0$, corresponds to the growing mode $\delta_1$, resp. decaying 
mode $\delta_2$. Inspection of \eqref{f12} shows that $f < f_1$ if $\delta_1 \delta_2 > 0$ 
while $f<f_1$ or $f>f_1$ when $\delta_1 \delta_2<0$. 
For concreteness we consider the situation around today. If both modes are positive 
we have $f<f_1$ and hence $\g > \gamma_1$. As we go back in time ($\Omega_m \to 1$), $f\to f_2$ 
and so $\g \to \infty$ when $f\approx 0$.
Let us consider now the situation $\delta_1 \times \delta_2<0$ with $|\delta_1|>|\delta_2|$ 
today. Now we have today $f>f_1$ and therefore $\g < \gamma_1$. When we go backwards, 
$f\to \infty$ at some point where the absolute value of both modes become equal, hence in that 
case $\g \to -\infty$. These different situations are shown on figures \eqref{vecf} and 
\eqref{vecf2}. 

Another interesting point concerns the asymptotic future ($\Omega_m \to 0$). Inside GR, the 
solution to eq. \eqref{dg} gives $\g \to \gamma_{\infty}$, eq. \eqref{gasf}, with 
\be
f \propto \frac{1}{a^2H}\propto a^{-\frac12(1-3 w_{\infty})} \to 0~. \lb{finfty}
\ee  
The crucial point is that the same asymptotic behaviour is obtained for all cases where 
the decaying mode is present, up to a change of the prefactor in \eqref{finfty} 
which depends on the initial conditions and on the amplitude of the decaying mode with 
respect to the growing mode. 
This immediately follows from the fact that one can neglect the last term
in eq.\eqref{del} in this regime, so that
\begin{equation}
\delta_m = \delta_{\infty} + {\rm const} \int \, \frac{dt}{a^2} = \delta_{\infty} +
{\rm const} \int \, \frac{dN}{a^2H}~.
\end{equation}
The same occurs for the more general modified gravity eq.\eqref{delmod}, apart from
the case of an anomalously large growth of $G_{eff}$ in the future.
On the other hand we have in the future $(w_{\infty}<-\frac13)$
\be
\Omega_m \sim a^{3 w_{\infty}}~.
\ee
Using the definition of $\g$, eq. \eqref{Omgamma}, it is straightforward to obtain 
\be
\gamma = \frac{\ln f}{\ln\Omega_m } \to \gamma_{\infty} ~, \lb{gamas} 
\ee
for \emph{all} curves with $\gamma$ today smaller or larger than the value obtained when we 
integrate eq.\eqref{dg} with $\g(1)= \gamma_{-\infty} (=\frac{6}{11}$ for $w_{DE}=-1$). 
This is somehow complementary to the results obtained in \cite{LP18} where the limit of 
a vanishing decaying mode was considered, however allowing for models beyond GR. Of 
course, in standard cosmological scenarios, the decaying mode is negligible already deep 
in the matter era so this result is essentially of mathematical interest.
This is nicely illustrated with figure \eqref{vecf2} for $w_{DE}=-1$, all curves 
tend to the same limit $\gamma_{\infty}=\frac{2}{3}$ but only one curve, corresponding to 
the limit of a purely growing mode, tends to the finite value 
$\gamma_{-\infty}=\frac{6}{11}$ in the past. 

%\vfill\eject
%%%%%%%%%%%%%%%%%%%%%%%%%%%%%%%%%%%%%%%%%%%%%%%%%%%%%%%%%%%%%%%%%%%%%%%%%%%%%%%%%%%%%
%%%%%%%%%%%%%%%%%%%%%%%%%%%%%%%%%%%%%%%%%%%%%%%%%%%%%%%%%%%%%%%%%%%%%%%%%%%%%%%%%%%%%
%%%%%%%%%%%%%%%%%%%%%%%%%%%%%%%%%%%%%%%%%%%%%%%%%%%%%%%%%%%%%%%%%%%%%%%%%%%%%%%%%%%%

%%%%%%%%%%%%%%%%%%%%%%%%%%%%%%%%%%%%%%%%%%%%%%%%%%%%%%%%%%%%%%%%%%%%%%%%%%%%%%%%%%%
%%%%%%%%%%%%%%%%%%%%%%%%%%%%%%%%%%%%%%%%%%%%%%%%%%%%%%%%%%%%%%%%%%%%%%%%%%%%%%%%%%%
\section{Global evolution of the slope}
%%%%%%%%%%%%%%%%%%%%%%%%%%%%%%%%%%%%%%%%%%%%%%%%%%%%%%%%%%%%%%%%%%%%%%%%%%%%%%%%%%%
%%%%%%%%%%%%%%%%%%%%%%%%%%%%%%%%%%%%%%%%%%%%%%%%%%%%%%%%%%%%%%%%%%%%%%%%%%%%%%%%%%%
In this section we want to consider more closely the slope of $\g(\Omega_m)$ when $\g$ evolves 
with time and this will allow us to find an interesting connection with the constant $\g$ 
case. 
It is seen from \eqref{dg} that a solution can satisfy $\frac{d\gamma}{d\Omega_m}=0$ 
in the $(\gamma,\Omega_m)$ plane only on the curve $\Gamma(\Omega_m)$ defined as 
\be
\overline{w} = w_{DE}~,\lb{dg=0}
\ee
where $w_{DE}$ is the true equation of state (EoS) of the DE component of the system under 
consideration, while $\overline{w}$ is given in \eqref{wgcon}. 
%
%and would correspond to the EoS of DE if $\gamma$ were constant. 
%
We can view $\overline{w}(\Omega_m,\g)$ as a given function of both variables $\Omega_m$ and 
$\gamma$, while $w_{DE}$ is the EoS of DE given in function of $\Omega_m$. 
Hence in each value of $\Omega_m$, the curve $\Gamma(\Omega_m)$ takes that value of 
$\gamma$ such that $\overline{w} = w_{DE}$, namely 
\be
\overline{w}\left(\Omega_m,\Gamma(\Omega_m)\right) = w_{DE}(\Omega_m)~. \lb{dg=0b}
\ee
The fact that $\Gamma(\Omega_m)$ is not constant just means that the solution 
$\gamma(\Omega_m)$ of \eqref{dg} is not constant. 
  
\begin{figure}[h!]
   \begin{center}
   \includegraphics[height=8cm]{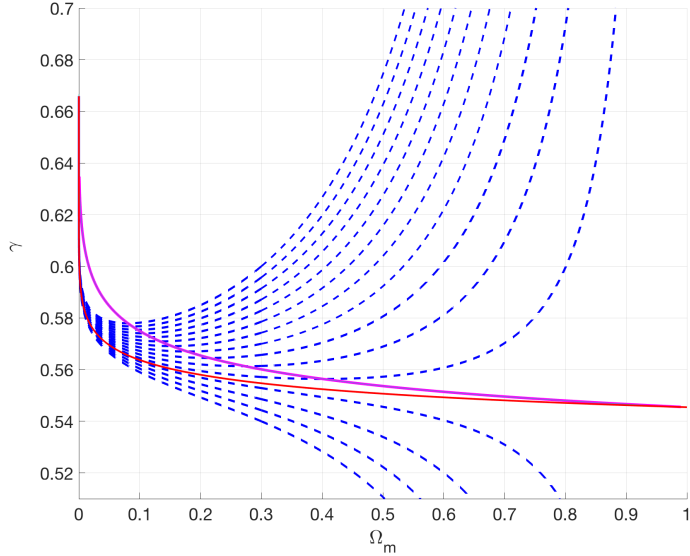}
   \caption{The curves $\g(\Omega_m)$ (red), with boundary condition 
    $\g(1)\equiv \g_{-\infty} = 6/11$, 
    and $\Gamma(\Omega_m)$ (purple), are shown for $w_{DE}=-1$. Above the curve 
    $\Gamma(\Omega_m)$, 
    $\frac{d\g}{d\Omega_m}>0$ while $\frac{d\g}{d\Omega_m}<0$ below it. 
    We see that $\Gamma(\Omega_m)$ is monotonically decreasing while the endpoints 
    $\gamma_{\infty}$ 
    and $\g_{-\infty}$, with $\gamma_{\infty} > \gamma_{-\infty}$, are identical for both curves. 
    Therefore $\g(\Omega_m)$ cannot cross $\Gamma(\Omega_m)$ at $0<\Omega_m<1$. The dashed 
    lines are integral 
    curves $\g(\Omega_m)$ however with $\g(1)\ne \g_{-\infty}= 6/11$. Still, as explained in 
    the text 
    while these curves diverge in the past, they all converge to the same limit 
    $\gamma_{\infty}$ in the future. Physically, all the dashed curves shown here correspond to 
    density perturbations with a non negligible decaying mode. Note that those dashed curves 
    which cross the curve $\Gamma(\Omega_m)$ satisfy $\frac{d\g}{d\Omega_m}=0$ where they 
    cross. 
   \label{vecf2} }
   %\label{vecf}
   \end{center}
\end{figure}

The curve $\Gamma(\Omega_m)$ defined by \eqref{dg=0} satisfies by construction in the 
asymptotic future
\be
\overline{w}\left(0,\Gamma(0)\right) = w_{\infty}~.
\ee
Hence we have in view of equalities \eqref{gasf} and \eqref{gconstasf}
\be
\Gamma(0) = \frac{3 w_{\infty} - 1}{6 w_{\infty}} = \gamma_{\infty}~. \lb{G0gas}
\ee
Analogously, we get in the asymptotic past 
\be
\overline{w}\left(1,\Gamma(1)\right) = w_{-\infty}~,
\ee
with, using \eqref{gasp} and \eqref{gconstasp}  
\be
\Gamma(1) = \frac{3(1 - w_{-\infty})}{5 - 6 w_{-\infty}} = \gamma_{-\infty}~.\lb{G1gas}
\ee
It is exactly here that we can use the crucial property reminded at the end of Section 2. 
Indeed, as it was shown in \cite{PSG16}, the solution $\gamma(\Omega_m)$ of \eqref{dg}  
starts at $\gamma_{-\infty}$ in the past and tends to $\gamma_{\infty}$ in the future (whence our 
choice of notation). 
In other words the curves $\Gamma(\Omega_m)$ and $\gamma(\Omega_m)$ meet at their endpoints 
(see figure \eqref{vecf2}). 

%%%%%%%%%%%%%%%%%%%%%%%%%%%%%%%%%%%%%%%%%%%%%%%%%%%%%%%
%%%%%%%%%%%%%%%%%%%%%%%%%%%%%%%%%%%%%%%%%%%%%%%%%%%%%%%

It is also seen from \eqref{dg} that all points $(\gamma,\Omega_m)$ above the 
curve $\Gamma(\Omega_m)$ satisfy $\frac{d\gamma}{d\Omega_m}>0$; on the other hand in the region 
below $\Gamma(\Omega_m)$ we have $\frac{d\gamma}{d\Omega_m}<0$. 
Let us assume that \eqref{dg=0} represents a monotonically decreasing function of 
$\Omega_m$, hence $\gamma_{\infty} > \gamma_{-\infty}$. This holds for example for all 
constant EoS of DE inside GR. 
Then it is clear that $\gamma(\Omega_m)$ cannot cross the 
curve $\Gamma(\Omega_m)$ at some value $\gamma_1$ with 
$\gamma_{-\infty} < \gamma_1 < \gamma_{\infty}$. 
If it did, $\gamma(\Omega_m)$ would cross with a zero derivative and have its 
minimum at $\gamma_1$. 
This is in contradiction with $\gamma_{-\infty} < \gamma_1$. 

This also shows that a change in the sign of the slope of $\gamma(\Omega_m)$ is possible 
only if 
$\Gamma(\Omega_m)$ is not a monotonically decreasing function of $\Omega_m$. 
Note that the slope of $\gamma(\Omega_m)$ does not vanish at $\Omega_m=1$ and $\Omega_m=0$ even 
though $\gamma(\Omega_m)$ and $\Gamma(\Omega_m)$ meet there. 
Actually it even diverges in $\Omega_m=0$ \cite{PSG16}.
Indeed, we have that the prefactor of the first term of \eqref{dg} vanish there leaving  
the slope a priori unspecified.   

For a given growth function $f$, a lower $\Omega_m$ implies a lower $\g$. For a 
given $\Omega_m$ on the other hand, a decrease in $f$ induces an increase in $\g$. All this is 
clearly understood from the equality $\gamma = \frac{\ln f}{\ln \Omega_m}$.
When $\gamma(\Omega_m)$ is monotonically descreasing, and this is the case for generic 
$w_{DE}$ inside GR, it is the second effect which prevails.  

%\vfill\eject
%%%%%%%%%%%%%%%%%%%%%%%%%%%%%%%%%%%%%%%%%%%%%%%%%%%%%%%%%%%%%%%%%%%%%%%%%%%%%%%%%%%%%%%%%
%%%%%%%%%%%%%%%%%%%%%%%%%%%%%%%%%%%%%%%%%%%%%%%%%%%%%%%%%%%%%%%%%%%%%%%%%%%%%%%%%%%%%%%%%
%%%%%%%%%%%%%%%%%%%%%%%%%%%%%%%%%%%%%%%%%%%%%%%%%%%%%%%%%%%%%%%%%%%%%%%%%%%%%%%%%%%%%%%%%

%%%%%%%%%%%%%%%%%%%%%%%%%%%%%%%%%%%%%%%%%%%%%%%%%%%%%%%%%%%%%%%%%%%%%%%%%%%%%%%%%%%%%%%%%
%%%%%%%%%%%%%%%%%%%%%%%%%%%%%%%%%%%%%%%%%%%%%%%%%%%%%%%%%%%%%%%%%%%%%%%%%%%%%%%%%%%%%%%%%
\section{Systems with a non monotonic $\g(\Omega_m)$}
%%%%%%%%%%%%%%%%%%%%%%%%%%%%%%%%%%%%%%%%%%%%%%%%%%%%%%%%%%%%%%%%%%%%%%%%%%%%%%%%%%%%%%%%
%%%%%%%%%%%%%%%%%%%%%%%%%%%%%%%%%%%%%%%%%%%%%%%%%%%%%%%%%%%%%%%%%%%%%%%%%%%%%%%%%%%%%%%%
We can consider now several cases where the growth index is not a monotonically decreasing 
function of $\Omega_m$. 

%%%%%%%%%%%%%%%%%%%%%%%%%%%%%%%%%%%%%%%%%%%%%%
\subsection{Varying equation of state $w_{DE}$}
%%%%%%%%%%%%%%%%%%%%%%%%%%%%%%%%%%%%%%%%%%%%%%
The first case that comes to one's mind is an oscillating equation of state $w_{DE}$. Indeed, 
it is easy to show that $\Gamma(\Omega_m)$ is no longer monotonically decreasing in this 
case and that $\gamma$ may oscillate provided the oscillations in $w_{DE}$, and hence in 
$\Gamma(\Omega_m)$, are sufficiently pronounced. Observations however do not seem to favour 
such oscillations. We rather want to investigate whether a smoothly varying EoS can lead 
to a change in slope of $\gamma$. For this purpose we use the parametrization in terms 
of $x\equiv \frac{a}{a_0}$ \cite{CP01}
\be
w_{DE}= w_0 + w_a(1-x)~.   \lb{CPL}
\ee
We consider the cases with 
\be
w_{-\infty}=w_0 + w_a < 0~~~~~~~~~~~~~~~~~~w_a>0~.
\ee
The first inequality is to enforce the domination of matter (and the disppearance of DE) 
in the past. We also restrict our attention to $w_a>0$ in order to have a varying EoS leading 
to the opposite case in the future, namely DE domination and disappearance of matter. 
Also we will not impose the restriction $w_{DE}\ge -1$ valid for quintessence 
(a self-interacting scalar field minimally coupled to gravity) and admit arbitrarily negative 
values of $w_{DE}$.
%%%%%%%%%%%%%%%%%%%%%%%%%%%%%%%%%%%%%%%%%%%%%%%%%%
\begin{figure}[h!]
   \begin{center}
   \includegraphics[height=8cm]{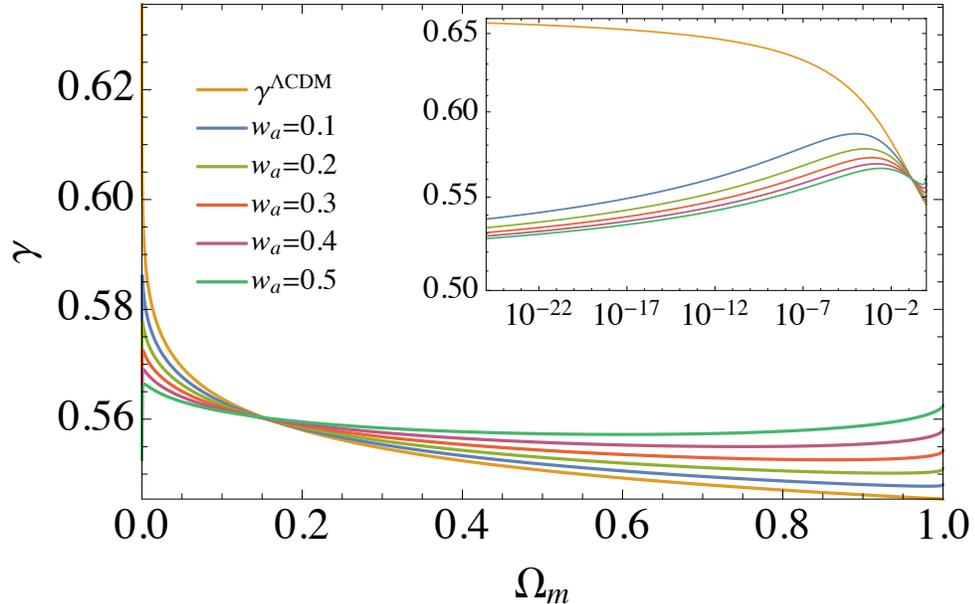}
   \caption{The evolution of $\g$ is shown for CPL varying $w_{DE}$. We see that $\g(\Omega_m)$ 
   is 
   essentially decreasing except in the remote future and, for $w_a\gtrsim 0.2$, in the far 
   past. Note the value $\gamma_{\infty} = \frac{1}{2}$ as expected for $w_{\infty} = -\infty$.} 
   \label{figCPL} 
   %\label{vecf}
   \end{center}
\end{figure}
%%%%%%%%%%%%%%%%%%%%%%%%%%%%%%%%%%%%%%%%%%%%%%%%%
We consider conservative values $0<w_a\le 0.5$ and $w_0\approx -1$.
We obtain that $\gamma(x)$ is quasi-constant deep in the matter dominated stage and starts 
decreasing (with the expansion) in the past typically at $z\simeq 10$, this decrease being 
stronger as $w_a$ is larger. In the future, a bump is obtained typically at $z\simeq -0.9$, 
which is higher for lower $w_a$, this bump would essentially disappear for $w_a=0.8$. 
Finally in the asymptotic future $\gamma \to \frac{1}{2}$ as we expect for 
$w_{\infty}\to -\infty$. 

In function of $\Omega_m$, $\gamma$ will be increasing deep in the DE 
domination typically till $\Omega_m\simeq 10^{-5}$, then decreasing till $\Omega_m\simeq 0.7$, 
and finally increasing again for $w_a\gtrsim 0.2$. To summarize, $\gamma(\Omega_m)$ is 
essentially 
slowly varying and decreasing from the past to the future in this case covering in particular 
the range probed by observations. The significant changes in slope are pushed around 
$\Omega_m\approx 0$. 
While this is interesting in itself, from an observational point of view models with 
$w_a\lesssim 0.2$ follow essentially the $\Lambda$CDM phenomenology regarding their growth 
index. Strong departure from $\Lambda$CDM takes place in the remote future only. 
These features are shown on figure \ref{figCPL}.

%%%%%%%%%%%%%%%%%%%%%%%%%%%%%%%%%%%%%%%%%%%%%%
\subsection{Beyond GR: a bump or a dip in $g$}
%%%%%%%%%%%%%%%%%%%%%%%%%%%%%%%%%%%%%%%%%%%%%%
There is however another way to have a change of slope of 
$\gamma$, and possibly in the range probed by observations. This is in the framework of 
modified gravity (see e.g. \cite{MG}) when $g$ is substantially varying and this can take 
place even when $w_{DE}$ is almost constant and close to $-1$.
We have to use the modified version of \eqref{dg} with $F(\Omega_m;\g)$ replaced by 
$F(\Omega_m;g,\g)$ where $g$ can be some arbitrary function, no longer equal to one as in GR. 
As we have already said earlier, a decreasing $\Omega_m$ tends to decrease $\g$ while a 
decreasing $f$ tends to increase it, see eq.\eqref{gamas}. 
A substantially varying $g$ can work in both ways depending on how it boosts or dampens 
the growth and on the time this process takes place. In other words it can strongly affect 
$f$ even if the background evolution remains ``standard''.   
\vskip 5pt
\par\noindent
a) {\it Bump in $g$}: A boost in $g$ was already found some time ago for $f(R)$ modified 
gravity. Let us consider more closely this case. When the bump 
in $g$ has reached its maximum ($\approx \frac{4}{3}$) in the past, $g$ is already starting 
on low redshifts its decrease towards its asymptotic value $g_{\infty}=1$ while still being 
higher today than one. This explains the low value $\gamma_0\approx 0.42$ obtained in 
\cite{GMP08,MSY10} but also the large \emph{negative} slope 
$\frac{d\gamma}{dz}\approx -0.25$. It would also be possible in principle in $f(R)$ models to 
have a bump whose maximum is reached in the future 
with $g$ today already substantially larger than one. In that case $\gamma_0$ could be again 
substantially lower than $0.55$ however now the slope on low redshifts would be positive, 
increasing with $z$. 
This provides a good example for which a global analysis gives more insight: $\g$ today lower 
than $\gamma^{\Lambda CDM}$ with $\frac{d\g}{dz}<0$ occurs if we have presently passed a bump in 
$g$.

\vskip 5pt
\par\noindent
b) {\it Dip in $g$}: While in $f(R)$ models we have $g\ge 1$, one can have modified gravity 
models with gravity weaker than GR. Then a decrease of $g$ on low redshifts yields an 
increasing $\g$ on these redshifts. This is in agreement with results found in \cite{GKPP18}.   
In this case too, a global analysis gives more insight: on a background 
with $w_{DE}=-1$, $\g$ today higher than $\gamma^{\Lambda CDM}$ with $\frac{d\g}{dz}>0$ tells us 
that we have presently passed a dip in $g$.
\vskip 5pt
\par
\begin{figure}[h!]
   \begin{center}
   \includegraphics[scale=0.5]{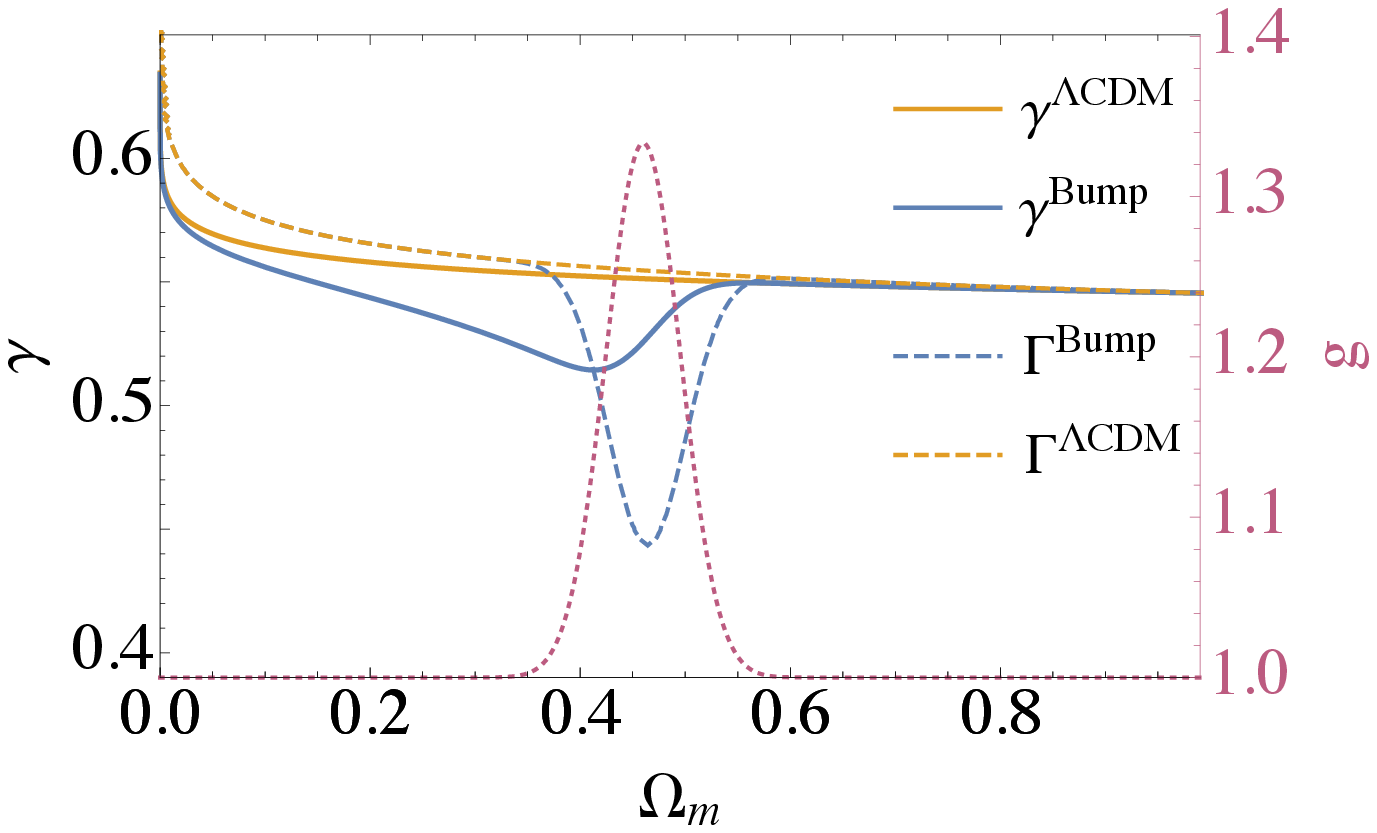}
   \includegraphics[scale=0.5]{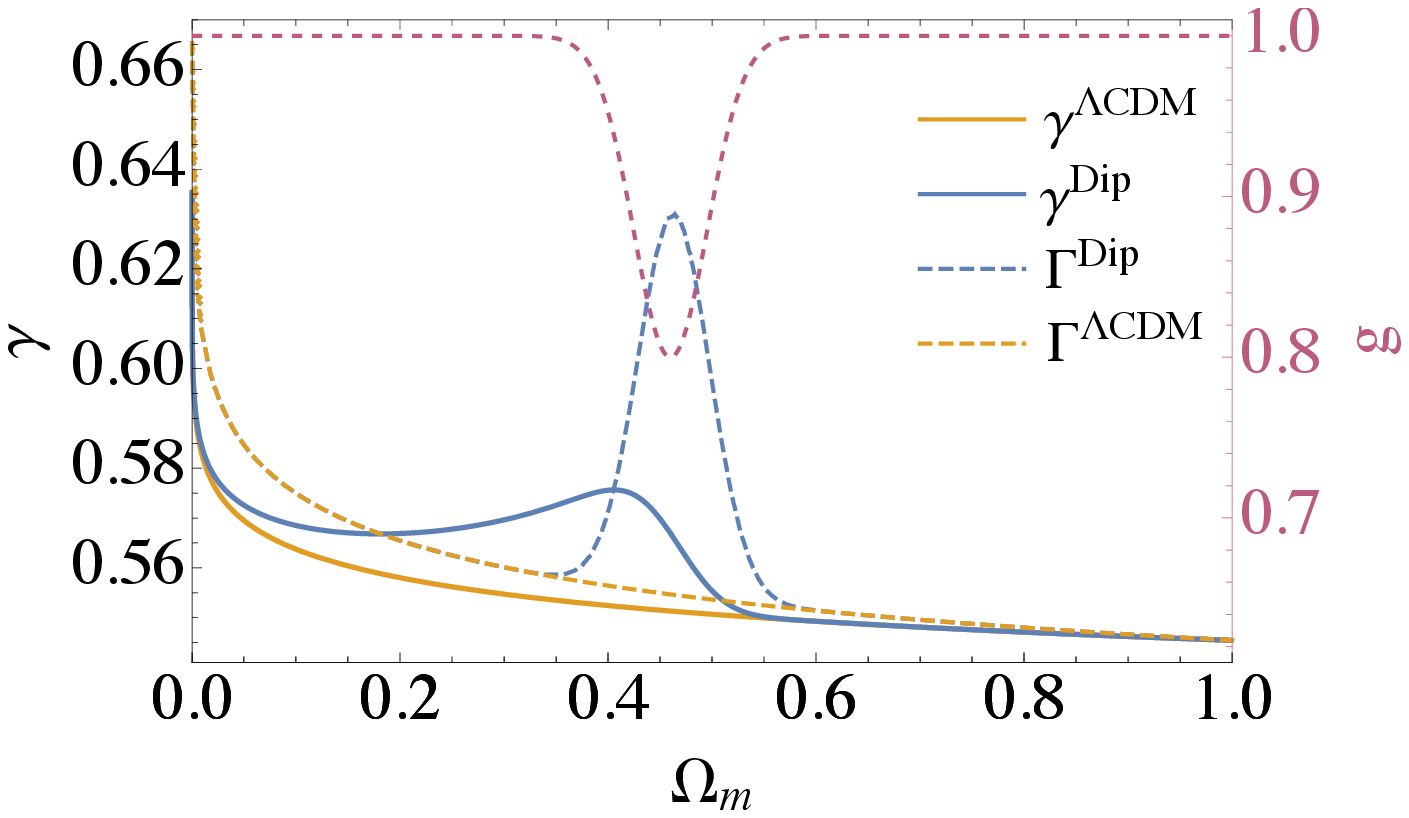}    
\caption{a) Left-hand panel: a schematic representation is given when there is a bump, here 
chosen to be slightly in the past (we take $\Omega_{m,0}=0.3$) while the background expansion is 
fixed ($w_{DE}=-1$). We see that $\g$ decreases first (with expansion) when $g$ goes up 
before increasing when $g$ decreases back to one. It is this last behaviour which was found 
today in many $f(R)$ models. Hence the sign of the derivative tells us that we (just) passed 
the maximum of the bump. Note that $\gamma^{bump}$ and $\gamma^{\Lambda CDM}$ shown here have the 
same asymptotic values both in past and future because $g\to 1$ in a smooth way and they have 
identical $w_{DE}=-1$. The behaviour on small redshifts and in the future captures essentially 
the $f(R)$ phenomenology. b) Right-hand panel: the same as on the left panel but now with a 
dip in $g$ starting today. At the present time, $\g>\gamma^{\Lambda CDM}$ and $\g$ is decreasing 
with expansion reflecting that $g$ is increasing and we have passed the minimum of the dip.} 
\label{fig-bump/dip} 
   \end{center}
\end{figure}

\noindent
Roughly speaking, we have $\g$ increasing, resp. decreasing when $g$ is decreasing, resp. 
increasing, so they have opposite behaviours.
We note finally a shift in the location of the extrema of $g$ and of $\g$ and as expected 
a dependence on the location and on the width of the bump in $g$.   
These properties are shown on figures \ref{fig-bump/dip}.
%%%%%%%%%%%%%%%%%%%%%%%%%%%%%%%%%%%%%%%%%%%%%%%
%%%%%%%%%%%%%%%%%%%%%%%%%%%%%%%%%%%%%%%%%%%%%%%
%%%%%%%%%%%%%%%%%%%%%%%%%%%%%%%%%
\subsection{Beyond GR: DGP model}
%%%%%%%%%%%%%%%%%%%%%%%%%%%%%%%%%
Another interesting example is provided by DGP brane models \cite{DGP00}. 
Let us note first that in this model both $g$ and $w_{DE}$ are given explicitly in 
terms of $\Omega_m$ as follows \cite{LSS04}
\be
w^{DGP} = - [1 +\Omega_m ]^{-1}~~~~~~~~~~~~~~~~~~~~g^{DGP} = 1 - 
                           \frac13 \frac{1 - \Omega_m^2}{1 + \Omega_m^2}~.
                                                                  \lb{wgDGP}
\ee
Hence it is straightforward to integrate $\g(\Omega_m)$ using the integration variable 
$\Omega_m$ as in 
\eqref{dg}. It is seen from \eqref{wgDGP} that $w^{DGP}$ tends to $w^{DGP}_{\infty}=-1$ in the 
future while $g^{DGP}$ tends to $g^{DGP}_{\infty}= \frac23$ in the future. So in the future the 
expansion looks like $\Lambda$CDM, however cosmic perturbations feel a weaker gravity. In 
the past $w^{DGP}$ tends to $w^{DGP}_{-\infty}=-\frac12$ while $g^{DGP}$ tends to 
$g^{DGP}_{-\infty}= 1$, so the gravitational driving force for the perturbations growth tends 
to its GR value. There are many interesting features when we solve for $\g(\Omega_m)$.  
%%%%%%%%%%%%%%%%%%%%%%%%%%%%%%%%%%%%%%%%%%%%%%%%%
\begin{figure}[h!]
   \begin{center}
   \includegraphics[scale=0.50]{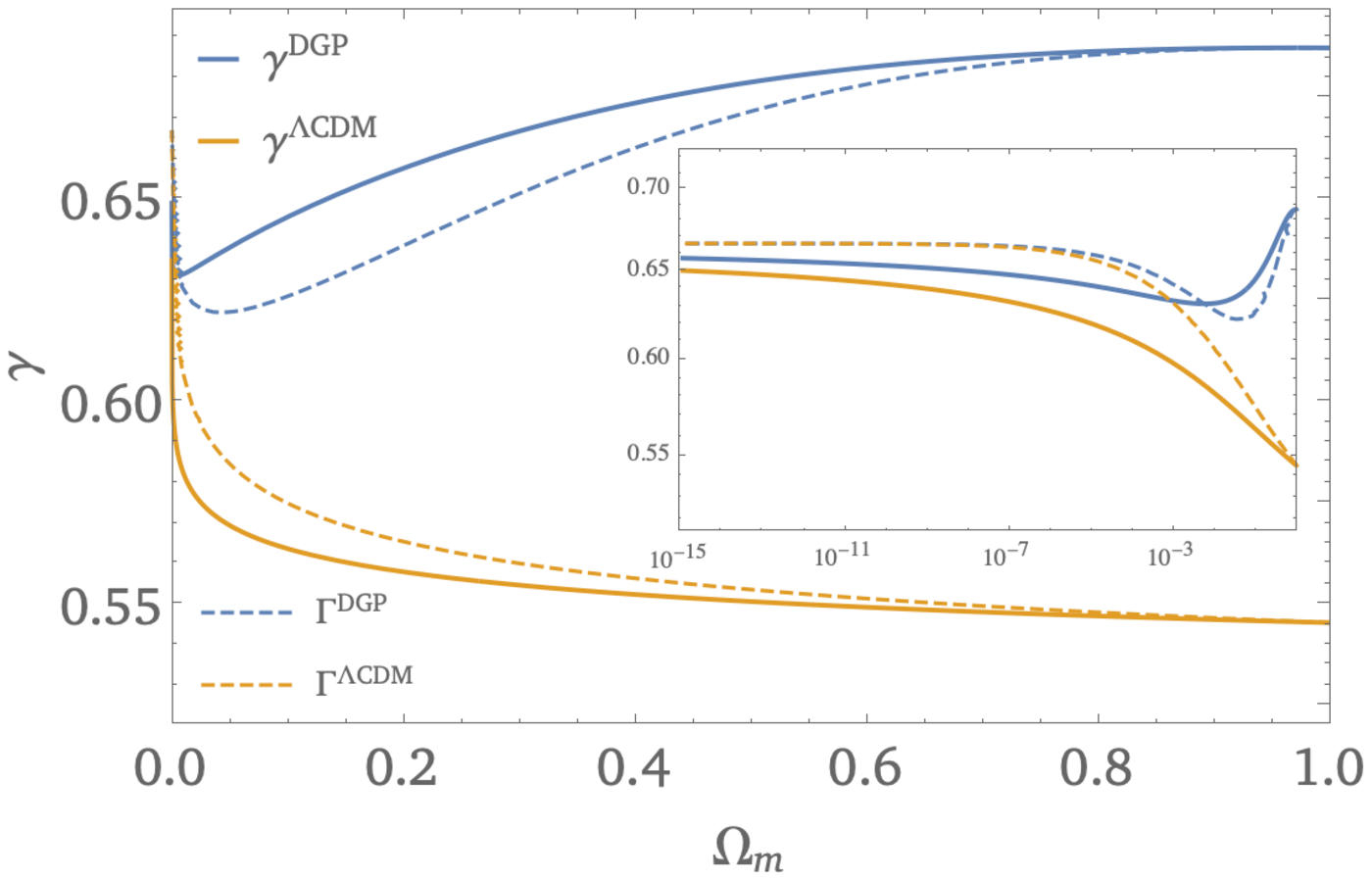}~~
   \includegraphics[scale=0.55]{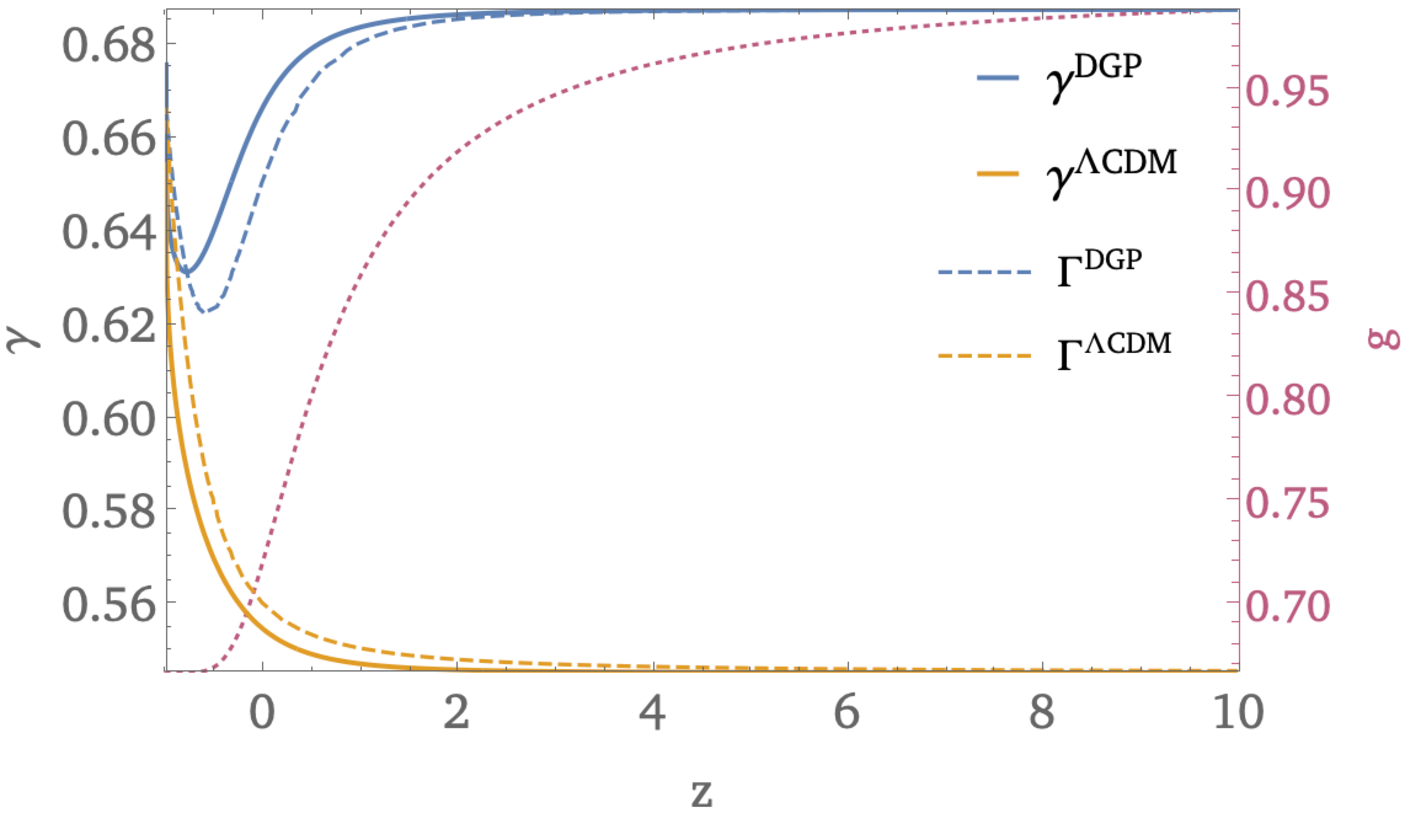}
   \caption{a)Left-hand panel: The behaviour of $\g$ in DGP models is shown and compared 
to that in $\Lambda$CDM. We see that $\g^{DGP}> \Gamma^{DGP}$ and hence $\g^{DGP}(\Omega_m)$ 
is an 
increasing function, except for the tiny interval $\Omega_m \lesssim 10^{-3}$. While 
$w_{-\infty}^{DGP}=-\frac12$, $\g_{-\infty}^{DGP} = \frac{11}{16}$ and not $\frac{9}{16}$ as one 
would have in GR. This is because the DGP model does not tend to GR in a smooth enough way, 
namely $\left(\frac{dg^{DGP}}{d\Omega_m}\right)_{-\infty} = \frac13\ne 0$. In the future 
however, while DGP does not tend to GR with $g^{DGP}_{\infty}= \frac23$ and $w^{DGP}_{\infty}=-1$, 
the same asymptotic value $\g_{\infty}^{DGP} = \frac23$ is obtained as in GR for $w_{\infty}=-1$ 
because gravity in DGP is weaker than in GR in the far future. Note that all four curves tend 
to $\frac23$ in the future. 
b) Right-hand panel; The same in function of $z$. We see 
that a value very close to $\frac{11}{16}$ is quickly reached in the past. As 
$\frac{11}{16}>\frac23$, $\Gamma^{DGP}(\Omega_m)$ could be monotonically increasing, which it is 
except for the tiny range $\Omega_m \lesssim 10^{-3}$ where $\g^{DGP}(\Omega_m)$ is a decreasing 
function.}
\label{fig-DGP} 
   \end{center}
\end{figure} 
%%%%%%%%%%%%%%%%%%%%%%%%%%%%%%%%%%%%%%%%%%%%%%%%%
Let us start with the behaviour in the future. As gravity becomes \emph{weaker} than GR in the 
future, $\g^{DGP}$ tends to $\g_{\infty}^{GR}$ corresponding to $w_{\infty}^{GR}= w_{\infty}^{DGP}=-1$. 
Note that this can still be so even if $g$ grows but not too quickly, see the discussion 
in \cite{LP18}.
This can also be understood by looking at the asymptotic form of equation 
\eqref{dg} for $\Omega_m \to 0$. Indeed, the same equation is obtained in GR when $g\to$ 
constant, also for $g\ne 1$, and even more generally if $g \Omega_m^{1-\g}\to 0$. 
Hence we obtain
\be
\g_{\infty}^{DGP} = \frac23~. \lb{ginfDGP}
\ee
In other words the same future asymptotic value is obtained as for $\Lambda$CDM. 

We turn now our attention to the behaviour in the past.
Curiously, while $g^{DGP}_{-\infty}= 1$ we would naively expect to have the same relation 
between $\g_{-\infty}^{DGP}$ and $w^{DGP}_{-\infty}=-\frac12$ as between 
$\g_{-\infty}^{GR}$ and $w^{GR}_{-\infty}=-\frac12$. However this is not so due to the non-vanishing 
derivative 
\be
\left(\frac{dg^{DGP}}{d\Omega_m}\right)_{-\infty} = \frac13~.
\ee
In order to find $\g(\Omega_m)$, we can use the same method as in \cite{PSG16} and write 
eq.\eqref{dg} for $\Omega_m \to 1$. After some calculation, solving for $\g(\Omega_m)$ we 
obtain in the 
asymptotic past for the finite constant solution $\g_{-\infty}$ for arbitrary modified 
gravity models 
\be
\g_{-\infty} = \frac{3 \left( w_{-\infty} - 1 - d \right)}
                                   {6 w_{-\infty} - 5}~, \lb{gpastmod}
\ee
where we have set 
\be
d\equiv \left(\frac{dg}{d\Omega_m}\right)_{-\infty}~.
\ee
The GR result is recovered setting $d=0$. Note that a similar relation was found between 
a constant $\g$ corresponding to $\overline{w}_{-\infty}= w_{-\infty}$ in an arbitrary modified 
gravity model \cite{PSG16}. We finally obtain for the DGP model 
\be
\g_{-\infty}^{DGP} = \frac{11}{16}~.
\ee 
So the (finite) $\g_{-\infty}$ corresponding to $w_{-\infty} = -\frac12$ has the value 
$\g_{-\infty}^{DGP} = \frac{11}{16}$ for a DGP model instead of $\frac{9}{16}$ for GR.
The derivation given here relies solely on the properties of the DGP model, eq. \eqref{wgDGP}, 
and exhibits the origin of this anomalous value in an explicit way 
(see \cite{LC07}, \cite{H14} for other approaches).

This value for $\g_{-\infty}^{DGP}$ corresponds to the constant $\g^{DGP}$ (and $g=g^{DGP}$) 
yielding identical equations of state, $\overline{w}_{-\infty}= w_{-\infty}^{DGP}$, in the 
past. We can summarize this in the following way
\be
\g_{-\infty}^{DGP}\left(w^{DGP}_{-\infty}\right) = \g \left(\overline{w}_{-\infty} = 
                         w_{-\infty}^{DGP}\right)~. \lb{eqpast}
\ee  

The left hand side corresponds to a varying $\g$ -- the true $\g$ obtained for $w^{DGP}$ and 
$g^{DGP}$ -- while the right hand side corresponds to a constant $\g$ in a modified gravity 
model with $g=g^{DGP}$.
A similar equality holds in the future and we have shown it here with the DGP model
\be
\g_{\infty}^{DGP}\left(w^{DGP}_{\infty}\right) = \g \left(\overline{w}_{\infty} = 
                         w_{\infty}^{DGP}\right)~. \lb{eqfuture}
\ee
Analogous equalities were found in \cite{PSG16} for GR and imply that the curves 
$\Gamma(\Omega_m)$ 
and $\g(\Omega_m)$ meet at $\Omega_m=0$ and $\Omega_m=1$, a property that we have used in 
the previous section. 
We have generalized this result to a class of modified gravity models including DGP models.   

As a corollary, we have the following important result: any modified gravity model with 
$g(1)=1$, i.e. tending to GR in the past, however in a smooth way satisfying $d = 0$, 
will have the same limit $\g^{GR}_{-\infty}(w_{-\infty})$ as in GR.  

As we can see on figure \eqref{fig-DGP}, the curve $\Gamma^{DGP}(\Omega_m)$ and 
$\g^{DGP}(\Omega_m)$ 
intersect indeed at $\Omega_m= 0$ and $\Omega_m=1$. In contrast to generic models inside GR, 
here we have  
$\gamma^{DGP}_{\infty}= \frac23 < \gamma^{DGP}_{-\infty}= \frac{11}{16}$. 
Hence $\Gamma(\Omega_m)$ cannot be monotonically decreasing, at best it would be monotonically 
increasing. If that were the case, $\g^{DGP}(\Omega_m)$ would be monotonically increasing as 
well, 
always lying above $\Gamma^{DGP}$. Actually this is what happens between 
$\Omega_m \simeq 10^{-3}$ 
and $\Omega_m = 1$. It is only in the far future, $\Omega_m \lesssim 10^{-3}$, that the slope of 
$\gamma^{DGP}(\Omega_m)$ is negative (i.e. $\gamma^{DGP}$ increases with expansion) and 
that the behaviour is similar to $\Lambda$CDM. 
 
%%%%%%%%%%%%%%%%%%%%%%%%%%%%%%%%%%%%%%%%%%%%%%%%%%%%%%%%%%%%%%%%%%%%%%%%%%%%%%%%%%%%%%%%%
%%%%%%%%%%%%%%%%%%%%%%%%%%%%%%%%%%%%%%%%%%%%%%%%%%%%%%%%%%%%%%%%%%%%%%%%%%%%%%%%%%%%%%%%%
%%%%%%%%%%%%%%%%%%%%%%%%%%%%%%%%%%%%%%%%%%%%%%%%%%%%%%%%%%%%%%%%%%%%%%%%%%%%%%%%%%%%%%%%%

%%%%%%%%%%%%%%%%%%%%%%%%%%%%%%%%
%%%%%%%%%%%%%%%%%%%%%%%%%%%%%%%%
\section{Summary and conclusion}
%%%%%%%%%%%%%%%%%%%%%%%%%%%%%%%%
%%%%%%%%%%%%%%%%%%%%%%%%%%%%%%%%
The growth index $\gamma$ allows to distinguish efficiently the phenomenology of dark 
energy (DE) models and has been used extensively for this purpose (see e.g. \cite{gamma}).
In this work we have performed an analysis of the growth index evolution from deep in the 
matter era till the asymptotic future. In this way global properties are exhibited. While from 
an observational point of view the main focus lies in the low redshift behaviour of the 
growth index $\g$ still, a global analysis yields some interesting insight and results. 
Some of the properties found here become transparent when we perform a global analysis 
of the growth index evolution. 
We have shown that when the growth index had a bump, resp. dip, in the recent past while the 
background evolution is similar to $\Lambda$CDM, today it is substantially lower, resp. 
larger, than $0.55$  with a negative, resp. positive, slope $\frac{d\g}{d\Omega_m}$ reflecting 
that the gravitational coupling $G_{\rm eff}$ of the underlying modified gravity model is already 
decreasing, resp. increasing, with the expansion. The behaviour with a bump is a schematic 
representation of many $f(R)$ models \cite{GMP08,MSY10}, the second case was considered in e.g. 
\cite{GKPP18}. 

Using results valid for a constant growth index, we suggest a condition giving the global 
sign of the slope $\frac{d\g}{d\Omega_m}$: when the curve $\Gamma(\Omega_m)$ introduced in 
Section 4 is monotonically decreasing then we have \emph{globally} $\frac{d\g}{d\Omega_m}<0$. 
This is the case in particular for models inside GR with a constant equation of state 
$w_{DE}=$ constant. Another interesting point concerns the value of the growth index for a 
given cosmological model at a given time. Actually the growth index $\g$ can take a range of 
values. What is really meant by the value of $\gamma(\Omega_m)$ which
corresponds to a given model is its value when the decaying mode of the perturbations mode 
tends to zero. We show that while the presence of a substantial decaying mode does not change 
the value of $\g$ in the asymptotic future, this leads (as expected) to a divergence in the 
asymptotic past. It is only in the limit of a vanishing decaying mode that $\g$ takes a finite 
value from 
$\g_{-\infty}$ in the asymptotic past ($\Omega_m\to 1$) -- $\frac{6}{11}$ for $\Lambda$CDM -- 
up to $\g_{\infty}$ in the asymptotic future ($\Omega_m\to 0)$ -- $\frac{2}{3}$ for $\Lambda$CDM. 
 
We have studied further the global behaviour of $\g$ for the DGP model. In contrast to generic 
models in GR, we have $\g_{-\infty}^{DGP}=\frac{11}{16}>\g_{\infty}^{DGP}=
\frac{2}{3}$ with $w_{-\infty}^{DGP}=-\frac{1}{2}$ and $w_{\infty}^{DGP}=-1$. While 
$g^{DGP}\to \frac23$ in the future, we have $\g_{\infty}^{DGP}=\g_{\infty}^{\Lambda CDM}$, so we get 
the same relation as in GR between $\g_{\infty}^{DGP}$ and $w_{\infty}^{DGP}=-1$. Interestingly, 
while $g^{DGP}\to 1$ in the past so this model tends to GR in the past, 
$\g_{-\infty}^{DGP}\ne \frac{9}{16}$, the value expected in GR for $w_{-\infty}=-\frac{1}{2}$. This 
is because the DGP model does not tend to GR in a way which is ``smooth enough''. Indeed, it 
satisfies $\frac{d g^{DGP}}{d\Omega_m}_{-\infty}=\frac13\ne 0$. As a corrolary we find that any 
modified gravity model which tends to GR in the past yields a $\g_{-\infty}$ which is the same 
function of $w_{-\infty}$ as in GR provided $\frac{d g^{DGP}}{d\Omega_m}_{-\infty}=0$. 
Finally we find that $\g^{DGP}(\Omega_m)$ is monotonically \emph{increasing} from 
the past until the far future ($\Omega_m\approx 10^{-3}$) where it crosses the curve 
$\Gamma(\Omega_m)$ in accordance with the condition mentioned above. 
The results presented in this work indicate that a measurement of $\g$ on a significant part 
of the expansion could give interesting constraints and a deeper insight into the physics 
governing the Universe dynamics.   

\section{Acknowledgements} 
A.A.S. was partly supported by the project number 0033-2019-0005 of the
Russian Ministry of Science and Higher Education.                                  
%%%%%%%%%%%%%%%%%%%%%%%%%%%%%%%%%%%%%%%%%%%%%%%%%%%%%%%%%%%%%%%%%%%%%%%%%%%%%%%%%%%%%%%%%%%%%
%%%%%%%%%%%%%%%%%%%%%%%%%%%%%%%%%%%%%%%%%%%%%%%%%%%%%%%%%%%%%%%%%%%%%%%%%%%%%%%%%%%%%%%%%%%%%

%%%%%%%%%%%%%%
%%%%%%%%%%%%%%

\begin{thebibliography}{99}

\bibitem{SS00} V.~Sahni and A.~A.~Starobinsky, Int. J. Mod. Phys. D {\bf 9}, 373 (2000);   
T. Padmanabhan, Phys. Rep. {\bf 380}, 235 (2003);
P.~J.~E.~Peebles and B.~Ratra, Rev. Mod. Phys. {\bf 75}, 559 (2003); 
E.~J.~Copeland, M.~Sami and S.~Tsujikawa, Int. J. Mod. Phys. D {\bf 15}, 1753 (2006); 
V.~Sahni and A.~A.~Starobinsky, Int. J. Mod. Phys. {\bf 15}, 2105 (2006);
 M.~Li, X.-D.~Li, S.~Wang and Y.~Wang, Commun. Theor. Phys. {\bf 56}, 525 (2011).

\bibitem{WMEHRR13} D.~H.~Weinberg, M.~J.~Mortonson, D.~J.~Eisenstein, C.~Hirata, A.~G.~Riess 
and E.~Rozo, Phys. Rept. {\bf 530}, 87 (2013); 
L. Amendola {\it et al.}, Living Rev. Rel. {\bf16}, 6 (2013); 
P. Bull {\it et al.}, Phys. Dark Univ. {\bf 12}, 56 (2016).

\bibitem{SSS14} V.~Sahni, A.~Shafieloo and A.~A.~Starobinsky, Astrophys. J. {\bf 793}, 
 L40 (2014).

\bibitem{P84} P.~J.~E.~Peebles, Astrophys. J. {\bf 284}, 439 (1984).

\bibitem{LLPR91} O.~Lahav, P.~B.~Lilje, J.~R.~Primack and M.~J.~Rees, MNRAS {\bf 251}, 128 
 (1991).

\bibitem{LC07} E.~V.~Linder and R.~N.~Cahn, Astropart. Phys. {\bf 28} 481 (2007). 

\bibitem{PG07} D. Polarski and R. Gannouji, Phys. Lett. B {\bf 660}, 439 (2008).

\bibitem{PSG16}  D.~Polarski, A.~A.~Starobinsky and H. Giacomini, JCAP {\bf 1612}, 037 (2016).

\bibitem{GMP08} R.~Gannouji, B.~Moraes and D.~Polarski, JCAP {\bf 0902}, 034 (2009).

\bibitem{MSY10} H.~Motohashi, A.~A.~Starobinsky and J.~Yokoyama, Progr. Theor. 
 Phys. {\bf 123}, 887 (2010).

\bibitem{WS98} L.~Wang and P.~J.~Steinhardt, Astrophys. J. {\bf 508}, 483 (1998).

\bibitem{Starobinsky:1998fr} A.~A.~Starobinsky, JETP Lett. {\bf 68}, 757 (1998).

\bibitem{LHuillier:2019imn} B.~L'Huillier, A.~Shafieloo, D.~Polarski and A.~A.~Starobinsky,
arXiv:1906.05991.

\bibitem{BEPS00} B.~Boisseau, G.~Esposito-Far\`ese, D.~Polarski and A.~A.~Starobinsky,
 Phys. Rev. Lett. {\bf 85}, 2236 (2000).

\bibitem{LP18} E. V. Linder, D. Polarski, Phys. Rev. D{\bf 99}, 2, 023503 (2019)

\bibitem{CP01} M. Chevallier and D. Polarski, Int. J. Mod. Phys. D{\bf 10}, 213 (2001);
 E. V. Linder, Phys. Rev. Lett. {\bf 90}, 091301 (2003).

\bibitem{MG} A.~De Felice and S.~Tsujikawa, Living Rev. Rel. {\bf 13}, 3 (2010);
 T.~Clifton, P.~G.~Ferreira, A.~Padilla and  C.~Skordis, Phys. Rept. {\bf 513}, 1 (2012); 
 A.~Joyce, B.~Jain, J.~Khoury and  M.~Trodden, Phys. Rept. {\bf 568}, 1 (2015);
 Mustapha Ishak, Living Rev. Rel. {\bf 22}, 1 (2019). 

\bibitem{GKPP18} R. Gannouji, L. Kazantzidis, L. Perivolaropoulos and D. Polarski,
Phys. Rev. D{\bf 98}, 10, 104044 (2018).

\bibitem{DGP00} G. Dvali, G. Gabadadze and M. Porrati, Phys.Lett. B{\bf 485}, 208 (2000).

\bibitem{LSS04} A. Lue, R. Scoccimarro and G. D. Starkman, Phys.Rev. D{\bf 69}, 124015 (2004)

\bibitem{H14} Q.-G. Huang, Eur. Phys. J. C{\bf 74}, 2964 (2014). 

\bibitem{gamma} 
 V. Acquaviva, A. Hajian, D. N. Spergel, S. Das, Phys. Rev. D{\bf 78} 043514 (2008);\\ 
 Hao Wei, Phys. Lett. {\bf B664} 1 (2008);\\
 S. Nesseris, L. Perivolaropoulos, Phys. Rev. D{\bf 77}, 023504 (2008);
 Yungui Gong, Phys.Rev. D{\bf 78}, 123010 (2008);\\ 
 Puxun Wu, Hong Wei Yu, Xiangyun Fu, JCAP 0906, 019 (2009);\\
 R. Bean, M. Tangmatitham, Phys. Rev. D{\bf 81}, 083534 (2010);\\
 Seokcheon Lee, Kin-Wang Ng, Phys. Lett. B{\bf 688}, 1 (2010);\\ 
 A. Bueno belloso, J. Garcia-Bellido, D. Sapone, JCAP 1110, 010 (2011);\\ 
 S. Nesseris, S. Basilakos, E.N. Saridakis, L. Perivolaropoulos, Phys. Rev. D{\bf 88} 103010 
 (2013);\\ 
 K. Bamba, Antonio Lopez-Revelles, R. Myrzakulov, S.D. Odintsov, L. Sebastiani, 
 Class. Quant. Grav. 30 015008 (2013);\\ 
 S. Basilakos, J. Sol\`a, Phys. Rev. D{\bf 92}, no.12, 123501 (2015);\\
 I. de Martino, M. De Laurentis, S. Capozziello, Universe 1, no.2, 123 (2015);\\
 J. N. Dossett, M. Ishak, D. Parkinson, T. M. Davis, Phys.Rev. D{\bf 92}, no.2, 
 023003 (2015);\\ 
 A. B. Mantz et al., Mon. Not. Roy. Astron. Soc. 446, 2205 (2015);\\ 
 Alberto Bailoni, Alessio Spurio Mancini, Luca Amendola, arXiv:1608.00458;\\
 Xiao-Wei Duan, Min Zhou, Tong-Jie Zhang, arXiv:1605.03947;\\
 B. Wang, E. Abdalla, F. Atrio-Barandela, D. Pavon, Rept. Prog. Phys. 79, no.9, 096901 
 (2016);\\ 
 N. Nazari-Pooya, M. Malekjani, F. Pace, D. Mohammad-Zadeh Jassur, 
 Mon. Not. Roy. Astron. Soc. 458, no.4, 3795 (2016);\\
 M. Malekjani, S. Basilakos, Z. Davari, A. Mehrabi, M. Rezaei, 
 Mon. Not. Roy. Astron. Soc. 464, 1192 (2017);\\
 R. Gannouji, D. Polarski, Phys.Rev. D{\bf 98}, no.8, 083533 (2018)



 


\end{thebibliography}
\end{document}